\documentclass[12pt,superscriptaddress]{iopart}
\usepackage{iopams}
\usepackage{graphicx}
\usepackage{mathtext}

\begin{document}

\newcommand {\eps} {\varepsilon}
\newcommand {\wD} {\widetilde{D}}
\newcommand {\ws} {\widetilde{\sigma}}
\newcommand {\wq} {\widetilde{q}}
\newcommand {\la} {\left\langle}
\newcommand {\ra} {\right\rangle}

\title{Diffusion of a passive scalar by convective flows under parametric disorder}

\author{Denis S Goldobin$^{1,2}$ and Elizaveta V Shklyaeva$^1$}
\address{$^1$Department of Theoretical Physics, Perm State University,
        15 Bukireva str., 614990, Perm, Russia}
\address{$^2$Department of Physics, University of Potsdam,
        Postfach 601553, D--14415 Potsdam, Germany}
\ead{Denis.Goldobin@gmail.com}

\begin{abstract}
We study transport of a weakly diffusive pollutant (a passive
scalar) by thermoconvective flow in a fluid-saturated horizontal
porous layer heated from below under frozen parametric disorder.
In the presence of disorder (random frozen inhomogeneities of the
heating or of macroscopic properties of the porous matrix),
spatially localized flow patterns appear below the convective
instability threshold of the system without disorder.
Thermoconvective flows crucially effect the transport of a
pollutant along the layer, especially when its molecular diffusion
is weak. The effective (or eddy) diffusivity also allows to
observe the transition from a set of localized currents to an
almost everywhere intense ``global'' flow. We present results of
numerical calculation of the effective diffusivity and discuss
them in the context of localization of fluid currents and the
transition to a ``global'' flow. Our numerical findings are in a
good agreement with the analytical theory we develop for the limit
of a small molecular diffusivity and sparse domains of localized
currents. Though the results are obtained for a specific physical
system, they are relevant for a broad variety of fluid dynamical
systems.
\end{abstract}

%Uncomment for PACS numbers title message
\pacs{05.40.-a,    % Fluctuation phenomena, random processes, noise, and Brownian motion
%      44.25.+f,    % Natural convection
      44.30.+v,    % Heat flow in porous media
      47.54.-r,    % Pattern selection; pattern formation
      72.15.Rn     % Localization effects (Anderson or weak localization)
}
% Keywords required only for MST, PB, PMB, PM, JOA, JOB?
\vspace{2pc}
\noindent{\it Special Issue}: Article preparation, IOP journals
% Uncomment for Submitted to journal title message
%\submitto{\it J.\ Stat.\ Mech.}
% Comment out if separate title page not required
\maketitle

%\section*{Introduction}

The effect of localization in spatially extended linear systems
subject to a frozen random spacial inhomogeneity of parameters is
known as Anderson localization (AL). AL has been first discovered
and discussed for quantum systems~\cite{Anderson-1958}. Later on,
investigations were extended to diverse branches of classical and
semiclassical physics: wave optics ({\it
e.g.},~\cite{Rossum-Nieuwenhuizen-1999}), acoustics ({\it
e.g.},~\cite{Maynard-2001}), {\it etc}. The phenomenon has been
comprehensively studied and well understood mathematically for the
Schr\"odinger equation and related mathematical models ({\it
e.g.},~\cite{Froehlich-Spencer-1984,Lifshitz-Gredeskul-Pastur-1988,Gredeskul-Kivshar-1992}).
Also, the role of nonlinearity in these models has been addressed
in the literature (for instance, destruction of AL by
nonlinearity~\cite{Pikovsky-Shepelyansky-2008,Gredeskul-Kivshar-1992}).

Being well studied for conservative media (or systems) the
localization phenomenon did not receive a comparable attention for
active/dissipative ones as, {\it e.g.}, in problems of thermal
convection or reaction-diffusion. The main reason is that the
physical interpretations of formal solutions to the Schr\"odinger
equation and governing equations for active/dissipative media are
essentially different and, therefore, the theory of AL may be
extended to the latter only under certain strong restrictions
(this statement is discussed in details in the end of the next
section). Nevertheless, effects similar to AL can be observed in
fluid dynamical systems (\cite{Goldobin-Shklyaeva-PRE-2008};
in~\cite{Hammele-Schuler-Zimmermann-2006} the effect of parametric
disorder on the excitation threshold in one-dimensional
Ginzburg--Landau equation has been studied, but without attention
to localization effects). In this paper, we study an example: the
problem where localized thermoconvective currents excited under
parametric disorder crucially influence the process of transport
of a passive scalar ({\it e.g.}, a pollutant).

The paper is organized as follows. In \sref{sec1} we formulate the
specific physical problem we deal with, introduce the relevant
mathematical model, and discuss physical background for the
problem.
\Sref{sec2} presents the results of a numerical simulation. In
\sref{sec3} we develop an analytical theory for a certain limit
case.
\Sref{concl} ends the paper with conclusions.

\section{Problem formulation and basic equations}\label{sec1}
The modified Kuramoto--Sivashinsky equation
\begin{equation}
\dot{\theta}(x,t)=-\left(\theta_{xxx}(x,t)
 +q(x)\,\theta_x(x,t)-(\theta_x(x,t))^3\right)_x
\label{eq1-01}
\end{equation}
\noindent
is relevant for a broad variety of active media where pattern
selection occurs. It governs two-dimensional (2D) large-scale
natural thermal convection in a horizontal fluid layer heated from
below~\cite{Knobloch-1990,Shtilman-Sivashinsky-1991} and holds
valid for a turbulent fluid~\cite{Aristov-Frick-1989}, a binary
mixture at small Lewis number~\cite{Schoepf-Zimmermann-1989-1993},
a porous layer saturated with a
fluid~\cite{Goldobin-Shklyaeva-BR-2008,Goldobin-Shklyaeva-PRE-2008},
{\it etc}.\footnote{In these fluid dynamical systems, except the
turbulent one~\cite{Aristov-Frick-1989}, the plates bounding the
layer should be nearly thermally insulating for a large-scale
convection to arise.} Specifically, in the problems mentioned,
temperature perturbations $\theta$ are almost uniform along the
vertical coordinate $z$ and obey~\eref{eq1-01}.

To argue for a general validity of~\eref{eq1-01}, let us note the
following. Basic laws in physics are the conservation ones. This
fact quite often results in final equations having the form
$\partial_t[\mbox{quantity}]+\nabla\!\cdot\![\mbox{flux of
quantity}]=0$. With such conservation laws either for systems with
the sign inversion symmetry of the fields, which is wide spread in
physics, or for description of a spatiotemporal modulation of an
oscillatory mode, the original Kuramoto--Sivashinsky equation
({\it e.g.}, see~\cite{Michelson-1986}) should be rewritten in the
form~\eref{eq1-01}. On these grounds, we claim
equation~\eref{eq1-01} to describe pattern formation in a broad
variety of physical systems.

In the following we restrict our consideration to the case of
convection in a porous medium; nevertheless, the most of results
may be easily extended to the other physical systems mentioned.
\Eref{eq1-01}
is already dimensionless and below we introduce all parameters and
variables in appropriate dimensionless forms.

\begin{figure}[!t]
\center{
\begin{tabular}{cc}
  \sf (a)&\hspace{-10mm}\includegraphics[width=0.90\textwidth]%
 {gold-shkl-08jstat-fig1a.eps}\\[5pt]
  \sf (b)&\hspace{-10mm}\includegraphics[width=0.90\textwidth]%
 {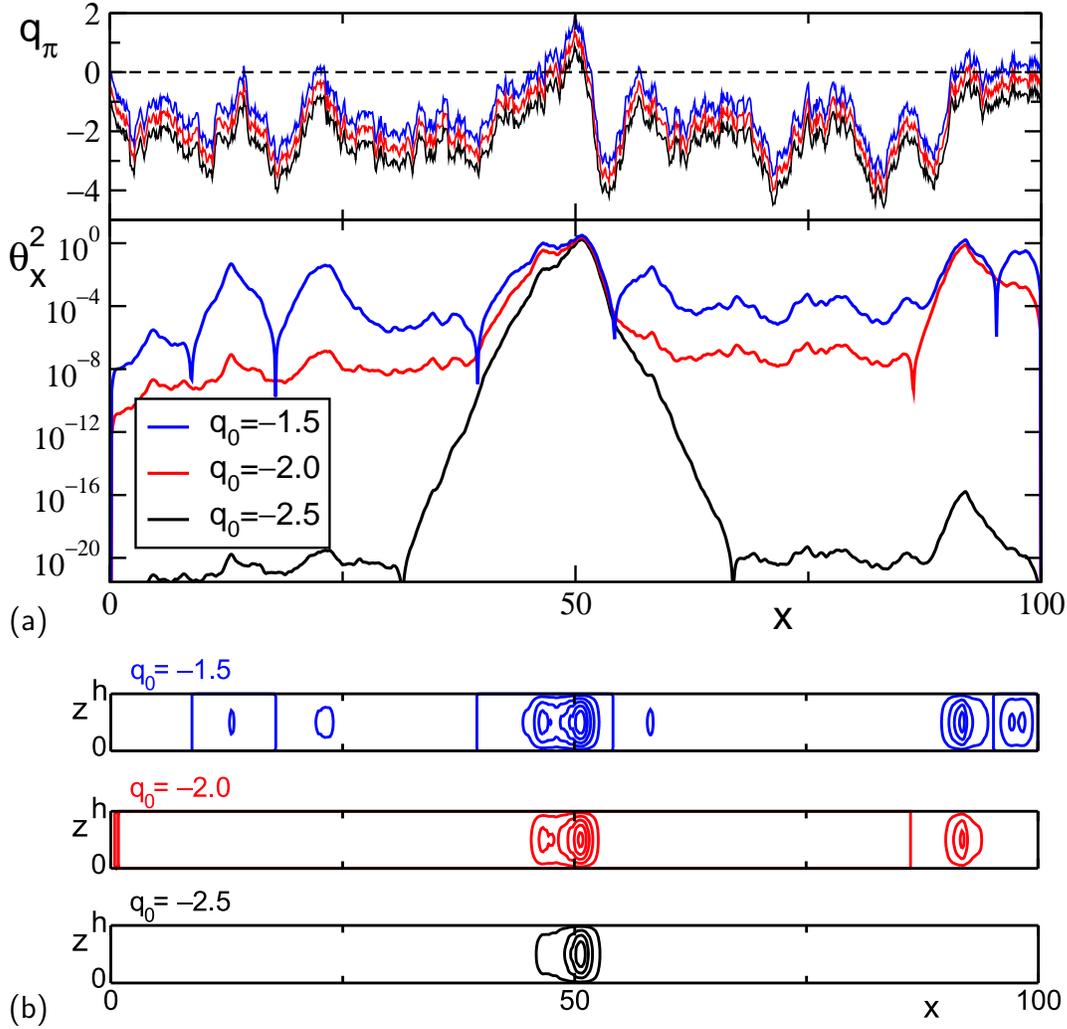}
\end{tabular}
}
  \caption{
(a):\,Establishing steady solutions to~\eref{eq1-01} for $q_0$
indicated in the plot are sets of exponentially localized patterns
[shown for one and the same realization of random inhomogeneity
$\xi(x)$ and $\eps=1$; $q(x)$ is represented by
$q_\pi(x)=\pi^{-1}\int_{x-\pi/2}^{x+\pi/2}q(x')\rmd x'$].
(b):\,The stream lines corresponding to the solutions in graph~(a)
are plotted for the case of convection in a porous layer
[cf.\,\eref{eq1-02}].}
  \label{fig1}
\end{figure}

Recall, the large-scale (or long-wavelength) approximation is
identical to the approximation of a thin layer and assumes that
the characteristic horizontal scales are large against the layer
height $h$. For large-scale convection, $(21/2)h^2q(x)$
(cf.\,\eref{eq1-01}, \cite{Goldobin-Shklyaeva-BR-2008}) represents
relative deviations of the heating intensity and of the
macroscopic properties of the porous matrix (porosity,
permeability, heat diffusivity, {\it etc.}) from the critical
values for the spatially homogeneous case. Thus, for positive
spatially uniform $q$, convection sets up, while for negative $q$,
all the temperature inhomogeneities decay. For convection in a
porous medium~\cite{Goldobin-Shklyaeva-BR-2008}, the macroscopic
fluid velocity field
\begin{equation}
\vec{v}=\frac{\partial\Psi}{\partial z}\vec{e}_x
 -\frac{\partial\Psi}{\partial x}\vec{e}_z\,,
 \qquad
\Psi=\frac{3\sqrt{35}}{h^3}\,z(h-z)\,\theta_x(x,t)\equiv f(z)\,\psi(x,t)\,,
\label{eq1-02}
\end{equation}
\noindent
where $\psi(x,t)\equiv\theta_x(x,t)$ is the stream function
amplitude, the reference frame is such that $z=0$ and $z=h$ are
the lower and upper boundaries of the layer, respectively
(\fref{fig1}b). Though the temperature perturbations
obey~\eref{eq1-01} for diverse convective systems, function
$f(z)$, which determines the relation between the flow pattern and
the temperature perturbation, is specific for each case.

%%%%%%%%%%%%%%%%%%%%%%%%%%%%%%%%%%%%%%%%%%%%%%%%%%%%%%%%%
%%%%%%%%%%%%%%%%%%%%%%%%%%%%%%%%%%%%%%%%%%%%%%%%%%%%%%%%%

Though \eref{eq1-01} is valid for a large-scale inhomogeneity
$q(x)$, which means $h|q_x|/|q|\ll1$, one may set such a hierarchy
of small parameters, namely $h\ll(h|q_x|/|q|)^2\ll1$, that a
frozen random inhomogeneity may be represented by white Gaussian
noise $\xi(x)$:
\[
q(x)=q_0+\xi(x),\quad \la\xi(x)\ra=0,\quad
\la\xi(x)\xi(x')\ra=2\eps^2\delta(x-x'),
\]
\noindent
where $\eps^2$ is the disorder intensity and $q_0$ is the mean
supercriticality ({\it i.e.}\ departure from the instability
threshold of the disorderless system). Numerical simulation
reveals only steady solutions to establish in~\eref{eq1-01} with
such $q(x)$~\cite{Goldobin-Shklyaeva-PRE-2008}.

Let us now discuss some general points related to the physical
problem under consideration. Obviously, the linearized form of
equation~\eref{eq1-01} in the stationary case, {\it i.e.},
\[
-\theta_{xxx}(x)-\xi(x)\,\theta_x(x)=q_0\,\theta_x(x)\,,
\]
is a stationary Schr\"odinger equation for $\psi=\theta_x$ with
$q_0$ instead of the state energy and $-\xi(x)$ instead of the
potential. Therefore, like for the Schr\"odinger equation ({\it
e.g.},
see~\cite{Froehlich-Spencer-1984,Lifshitz-Gredeskul-Pastur-1988,Gredeskul-Kivshar-1992}),
all the solutions $\psi(x)$ to the stationary linearized
equation~\eref{eq1-01} are spatially localized for arbitrary
$q_0$; asymptotically,
\[
\psi(x)\propto\exp(-\gamma|x|),
\]
where $\gamma$ is the localization exponent. Such a localization
can be easily seen for a solution to the nonlinear
problem~\eref{eq1-01} in
\fref{fig1}a for $q_0=-2.5$.

One should keep in mind, that, in the quantum Schr\"odinger
equation, localized modes are bound states of, {\it e.g.}, an
electron in a disordered media. Even the mutual nonlinear
interaction of these modes, which appears due to the
electron--electron interaction and leads to destruction of AL,
should be interpreted in the context of the specific physical
meaning of the quantum wave function. Therefore, the theory
developed for AL in quantum systems may not be directly extended
to active/dissipative media. Indeed, in~\eref{eq1-01}, all excited
localized modes of the linearized problem do mutually interact via
nonlinearity in a way where they irreversibly lose their identity
(unlike solitons in soliton bearing systems, which completely
recover after mutual collision). Thus, when the spatial density of
excited localized modes is large and these modes form an almost
everywhere intense flow, localization properties of formal
solutions to the linearized problem do absolutely not manifest
themselves.

Nevertheless, when excited modes are spatially sparse, solitary
exponentially localized patterns can be discriminated as reported
in~\cite{Goldobin-Shklyaeva-PRE-2008}. \Fref{fig1} shows sample
patterns for such a case. One can see that for negative $q_0$ the
spatial density of excited modes rapidly decreases as $q_0$
decreases and the pattern localization becomes more pronounced.
For a small spatial density of excited modes, one can distinguish
all these modes and introduce the observable quantifier $\nu$ of
the established steady pattern, which measures the spatial density
of the domains of excitation of convective flow; fortunately, an
empiric formula fits perfectly the numerically calculated
dependence~\cite{Goldobin-Shklyaeva-PRE-2008},
\begin{equation}
\nu\approx
\frac{1}{4\sqrt{1.95\,\pi}\,\eps^{2/3}|\wq_0|}
\exp\left(-\frac{1.95\,\wq_0^2}{4}\right),
\label{eq1-03}
\end{equation}
\noindent
where $\wq_0\equiv\eps^{-4/3}q_0$.

Here we would like to emphasize the fact of existence of
convective currents below the instability threshold of the
disorderless system. These currents may considerably and
nontrivially affect transport of a pollutant (or other passive
scalar), especially when its molecular diffusivity is small (for
instance, for microorganisms or suspensions the diffusion due to
Brownian motion is drastically weak against the possible
convective transport). Transport of a nearly indiffusive passive
scalar is the subject of our research, as a ``substance'' which is
essentially influenced by these localized currents and, thus,
provides an opportunity to observe manifestation of
disorder-induced phenomena discussed
in~\cite{Goldobin-Shklyaeva-PRE-2008}.

From the viewpoint of mathematical physics, there is one more
nontrivial question which is worthy to be addressed. In AL an
important topological effect takes place; while in 1D case all the
solutions are localized, in higher dimensions spatially unbounded
solutions appear ({\it e.g.},
\cite{Froehlich-Spencer-1984}). The modification of~\eref{eq1-01}
for the case of inhomogeneity in the both horizontal directions,
$q=q(x,y)$, ({\it e.g.}, see~\cite{Goldobin-Shklyaeva-BR-2008})
cannot be turned into the Schr\"odinger equation even after
linearization in the stationary case. Thus, there are no reasons
for any topological effects directly analogous to the one
mentioned for AL. Nevertheless, one may speak of a percolation
kind transition, where the domain of an intense convective flow
becomes globally connected for high enough $q_0$. Noteworthy, this
transition cannot be observed in 1D system~\eref{eq1-01} as there
is always a finite probability of a large domain of negative
$q(x)$ where the flow is damped and the domain of an intense flow
becomes disconnected. Essentially, the flow damped never decays
exactly to zero and, hence, one needs a formal quantitative
criterion for the absence of intense currents at a certain point.
On the other hand, this transition leads to a crucial enhancement
of transport of a nearly indiffusive scalar along the layer, and
the intensity of this transport can be used to detect the
transition immediately in the context that arises applied interest
to it. In this way, one also avoids introducing a formal
quantitative criterion. Remarkably, in the context of transport of
a passive scalar, that is the subject of the study we present, the
transition from a set of spatially localized currents to an almost
everywhere intense ``global'' flow can be observed in 1D
system~\eref{eq1-01} as well.

Let us describe the transport of a passive ({\it i.e.}, not
influencing the flow in contrast, for instance,
to~\cite{Goldobin-Lyubimov-2007}) pollutant by a steady convective
flow~\eref{eq1-02}. The flux $\vec{j}$ of the pollutant
concentration $C$ is
\begin{equation}
\vec{j}=\vec{v}\,C-D\nabla C\,,
\label{eq1-04}
\end{equation}
\noindent
where the first term describes the convective transport and the
second one represents the molecular diffusion, $D$ is the
molecular diffusivity. The establishing steady distributions of
the pollutant obey
\begin{equation}
\nabla\cdot\vec{j}=0\,.
\label{eq1-05}
\end{equation}
\noindent
\Eref{eq1-05}
yields a uniform along $z$ distribution of $C$ (see appendix),
\begin{equation}
\frac{\rmd C(x)}{\rmd x}=-\frac{J}{\displaystyle D+\frac{21\,\psi^2(x)}{2h^2D}}\,,
\label{eq1-06}
\end{equation}
\noindent
where $J$ is the constant pollutant flux along the layer. Note,
for the other convective systems we mentioned above, the result
differs only in the factor ahead of $\psi^2/D$.

\section{Effective diffusivity}\label{sec2}
In this section we introduce and consider the effective
diffusivity (for general ideas on introducing the effective
diffusivity one can consult, {\it
e.g.},~\cite{Frisch-1995,Majda-Kramer-1999}). Let us consider the
domain $x\in[0,L]$ with the imposed concentration difference
$\delta C$ at the ends. Then the establishing pollutant flux $J$
is defined by the integral [cf.\,\eref{eq1-06}],
\[
\delta C=-J\int\limits_0^L\frac{\rmd x}{\displaystyle D+\frac{21\,\psi^2(x)}{2h^2D}}\,.
\]
\noindent
For a lengthy domain the specific realization of $\xi(x)$ becomes
insignificant;
\[
\delta C=-J\,L\la\left(D+\frac{21\,\psi^2(x)}{2h^2D}\right)^{-1}\ra
\equiv-\sigma^{-1}J\,L\,,
\]
\noindent
Hence,
\[
J=-\sigma\frac{\delta C}{L},
\]
\noindent
{\it i.e.}\ $\sigma$ can be considered as an effective
diffusivity.

%%%%%%%%%%%%%%%%%%%%%%%%%%%%%%%%%%%%%%%%%%%%%%%%%%%%%%%%%
\begin{figure}[!t]
\center{
\includegraphics[width=0.89\textwidth]%
 {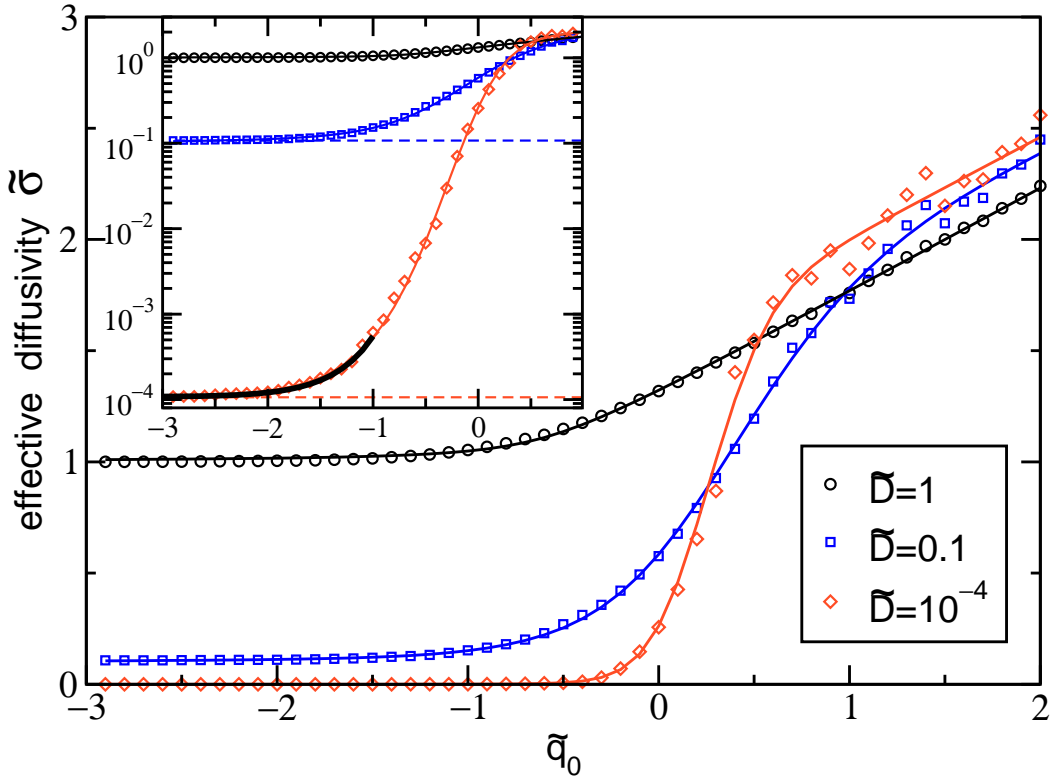}
}
  \caption{
Dependencies of effective diffusivity $\sigma$ on mean
supercriticality $q_0$ for molecular diffusivity $D$ indicated in
the plot. The bold black line in the inner plot represents the
analytical dependence (see \sref{sec3}).}
  \label{fig2}
\end{figure}
%%%%%%%%%%%%%%%%%%%%%%%%%%%%%%%%%%%%%%%%%%%%%%%%%%%%%%%%%

%%%%%%%%%%%%%%%%%%%%%%%%%%%%%%%%%%%%%%%%%%%%%%%%%%%%%%%%%
\begin{figure}[!t]
\center{
\includegraphics[width=0.67\textwidth]%
 {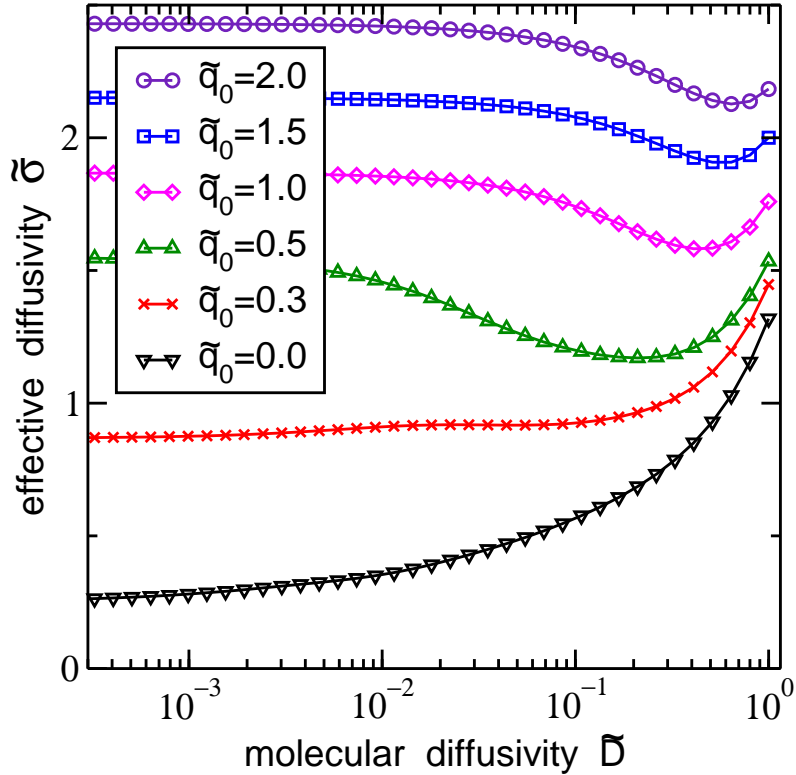}
}
  \caption{
Dependencies of the effective diffusivity on the molecular one for
nonnegative $\wq_0$}
  \label{fig3}
\end{figure}
%%%%%%%%%%%%%%%%%%%%%%%%%%%%%%%%%%%%%%%%%%%%%%%%%%%%%%%%%

The effective diffusivity
\begin{equation}
\sigma=\la\left(D+\frac{21\,\psi^2(x)}{2h^2D}\right)^{-1}\ra^{-1}
\label{eq2-01}
\end{equation}
\noindent
turns into $D$ for vanishing convective flow. For small $D$ the
regions of the layer, where the flow is damped, $\psi\ll1$, make
large contribution to the mean value appearing in~\eref{eq2-01}
and diminish $\sigma$, thus, leading to the locking of the
spreading of the pollutant.

Note, disorder strength $\eps^2$ can be excluded from equations by
the appropriate rescaling of parameters and fields. Thus, the
results on the effective diffusivity can be comprehensively
presented in the terms of $\wD$, $\ws$, and $\wq_0$:
\[
\wD=\sqrt{\frac{2}{21}}\,\eps^{4/3}hD,\qquad
\ws=\sqrt{\frac{2}{21}}\,\eps^{4/3}h\sigma,\qquad
\wq_0=\frac{q_0}{\eps^{4/3}}\,.
\]
\noindent
\Fref{fig2} provides calculated dependencies of effective
diffusivity $\ws$ on $\wq_0$ for moderate
and small values of molecular diffusivity $\wD$. Noteworthy,
\\
(i)\;for small $\wD$ a quite sharp transition of effective
diffusivity $\ws$ between moderate values and ones comparable with
$\wD$ occurs near $q_0=0$, suggesting the transition from an
almost everywhere intense ``global'' flow to a set of localized
currents to take place;
\\
(ii)\;below the instability threshold of the disorderless system,
where only sparse localized currents are excited, the effective
diffusion can be dramatically enhanced by these currents; {\it
e.g.}, for $\wD=10^{-4}$, $\wq_0=-1$, the effective diffusivity is
increased by one order of magnitude compared to the molecular
diffusivity.

\Fref{fig3}
shows dependencies of the effective diffusivity on the molecular
one for nonnegative $q_0$. Remarkably, for $\wq_0\gtrsim0.3$, the
dependencies possess a minimum which is in agreement with known
general results on interference between turbulent and molecular
diffusion~\cite{Saffman-1960-Mazzino-Vergassola-1997}. The reason
is the fact, that for the convective transport the molecular
diffusion plays a destructive role. Hence, for high $q_0$, where
convective flows are intense, the enhancement of the convective
transport prevails over the weakening of the diffusional one as
the molecular diffusivity tends to zero; on the contrary, for low
$q_0$, where convective flows are weak, the decrease of the
molecular diffusivity leads to the weakening of the transport.

\section{Analytical theory}\label{sec3}
Let us now analytically evaluate the effective diffusivity for a
small molecular one, $\wD\ll1$, and sparse domains of excitation
of convective flow, $\nu\ll1$. We have to calculate the average
\[
\beta\equiv\la\big(\wD+\psi^2(x)/\wD\big)^{-1}\ra
\]
\noindent
[cf.\ \eref{eq2-01}] which, due to ergodicity, can be evaluated
not only as an average over $x$ for a given realization of
$\xi(x)$, but also as an average over realizations of $\xi(x)$ at
a certain point $x_0$. Let us set the origin of the $x$-axis at
$x_0$. Hence,
$\beta=\langle\big(\wD+\psi^2(0)/\wD\big)^{-1}\rangle_\xi$.

%%%%%%%%%%%%%%%%%%%%%%%%%%%%%%%%%%%%%%%%%%%%%%%%%%%%%%%%%
\begin{figure}[!t]
\center{
\includegraphics[width=0.68\textwidth]%
 {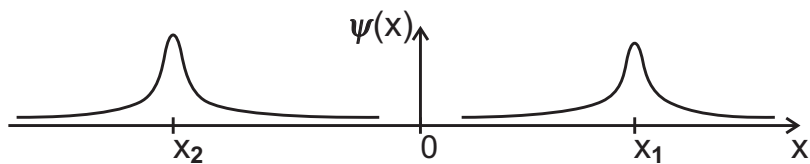}
}
  \caption{
Sketch of two localized flow patterns being nearest to the origin}
  \label{fig4}
\end{figure}
%%%%%%%%%%%%%%%%%%%%%%%%%%%%%%%%%%%%%%%%%%%%%%%%%%%%%%%%%

When the two nearest to the origin excitation domains are distant
and localized near $x_1>0$ and $x_2<0$ (see~\fref{fig4}),
\begin{equation}
\psi(0)\approx\psi_1\e^{-\gamma x_1}+\psi_2\e^{-\gamma|x_2|},
\label{eq3-01}
\end{equation}
\noindent
where $\psi_{1,2}$ characterize the amplitude of flows excited
around $x_{1,2}$. For small $\wD$ and density $\nu$, the
contribution of the excitation domains to $\beta$ is negligible
against the one of the regions of a weak flow. Therefore, we do
not have to be very accurate with the former and may utilize
expression~\eref{eq3-01} even for small $x_{1,2}$,
\begin{eqnarray}
\beta=\la\frac{1}{\wD+\psi^2(0)/\wD}\ra_\xi\nonumber\\
\qquad
 =\la\int\limits_0^{\infty}\rmd x_1\int\limits_0^{\infty}\rmd x_2
 \frac{p(x_1)\,p(x_2)}{\wD+\wD^{-1}(\psi_1\e^{-\gamma x_1}+\psi_2\e^{-\gamma x_2})^2}
 \ra_{\psi_1,\psi_2}
\label{eq3-02}
\end{eqnarray}
\noindent
where $p(x_1)$ [\,$p(x_2)$] is the density of the probability to
observe the nearest right [left] excitation domain at $+x_1$
[$-x_2$]. For probability distribution $P(x_1\!>\!x)$, one finds
 $P(x_1\!>\!x+\rmd x)=P(x_1\!>\!x)\,(1-\nu\rmd x)$,
{\it i.e.}, $\frac{\rmd}{\rmd x}P(x_1\!>\!x)=-\nu P(x_1\!>\!x)$.
Hence, $P(x_1\!>\!x)=\e^{-\nu x}$, and probability density
$p(x)=|\frac{\rmd}{\rmd x}P(x_1\!>\!x)|=\nu\e^{-\nu x}$. With
regard to averaging over $\psi_{1,2}$, it is important that the
multiplication of $\psi_{1,2}$ by factor $F$ is effectively
equivalent to the shift of the excitation domain by
$\gamma^{-1}\ln{F}$ which is insignificant for $F\sim1$ in the
limit case we consider. Hence, one can assume $\psi_{1,2}=\pm1$
(the topological difference between $\psi_1\psi_2<0$ and
$\psi_1\psi_2>0$ is not to be neglected) and rewrite~\eref{eq3-02}
as
\begin{eqnarray}
 \beta=\frac{1}{2}\int\limits_0^{\infty}\rmd x_1\int\limits_0^{\infty}\rmd x_2\,
 \nu^2\e^{-\nu(x_1+x_2)}
 \left[\frac{1}{\wD+\wD^{-1}(\e^{-\gamma x_1}+\e^{-\gamma x_2})^2}
 \right.\nonumber\\
 \hspace{70mm}\left.
 {}+\frac{1}{\wD+\wD^{-1}(\e^{-\gamma x_1}-\e^{-\gamma x_2})^2}\right].
 \nonumber
\end{eqnarray}
\noindent
These integrals can be evaluated for $\nu/\gamma\ll1$, and one
finds
\begin{equation}
\ws=\frac{1}{\beta}\approx\wD\left(\frac{2}{\wD}\right)^\frac{2\nu}{\gamma}\!\!.
\label{eq3-03}
\end{equation}
\noindent
For $\wq_0\!<\!-1$, one can use the asymptotic expressions for
$\nu$ [equation~\eref{eq1-03}] and
\[
\gamma=\eps^{-\frac{2}{3}}\left(|\wq_0|^\frac{1}{2}-\frac{1}{4}|\wq_0|^{-1}-\frac{5}{32}|\wq_0|^{-\frac{5}{2}}+\dots\right).
\]
\noindent
The latter expression is known from the classical theory of AL
(cf.\,\cite{Lifshitz-Gredeskul-Pastur-1988,Gredeskul-Kivshar-1992}).

In~\fref{fig2}, one can see analytic expression~\eref{eq3-03} to
match the numerically evaluated $\ws$ for $\wD=10^{-4}$,
$\wq_0\!<\!-1$ quite well. With~\eref{eq3-03}, one can evaluate
the convectional enhancement of the effective diffusivity below
the excitation threshold of the disorderless system, and it is
given by factor $\ws/\wD=(2/\wD)^{2\nu/\gamma}$ which can be large
for small $\wD$.

\section{Conclusion}\label{concl}
Summarizing, we have studied the transport of a pollutant in a
horizontal fluid layer by spatially localized 2D thermoconvective
currents appearing under frozen parametric disorder. Though the
specific physical system we have considered is a horizontal porous
layer saturated with a fluid and confined between two nearly
thermally insulating plates, our results can be trivially extended
to a broad variety of fluid dynamical systems (like ones studied
in~\cite{Knobloch-1990,Shtilman-Sivashinsky-1991,Aristov-Frick-1989,Schoepf-Zimmermann-1989-1993}).
We have calculated numerically the dependence of the effective
diffusivity on the molecular one and the mean supercriticality
(see figures~\ref{fig2},\,\ref{fig3}). In particular, for a nearly
indiffusive pollutant ($\wD\ll1$), first, we have observed the
transition from a set of localized flow patterns to an almost
everywhere intense ``global'' flow, which results in a soar of the
effective diffusivity from values comparable with the molecular
diffusivity up to moderate ones. Second, we have found convective
currents to considerably enhance the effective diffusivity even
below this transition. For the latter effect the analytical
theory, which perfectly describes the limit of $\wD\ll1$,
$\nu\ll1$, has been developed [equation~\eref{eq3-03}].

\ack{
The authors thank M.\,Zaks and S.\,Shklyaev for useful discussions
and O.\,Khlybov and A.\,Alabuzhev for technical support with
computational facilities.\footnote{Calculation of statistical
properties of states of an extensive distributed stochastic
system, like the one performed in this work, is extremely
CPU-time-consuming.} DG acknowledges the Foundation ``Perm
Hydrodynamics,'' the BRHE--program (CRDF Grant no.\,Y5--P--09--01
and MESRF Grant no.\,2.2.2.3.16038), and the VW--Stiftung for
financial support.}

\appendix
\setcounter{section}{1}
\section*{Appendix: Diffusion by stationary flow}
The following derivation of equation~\eref{eq1-06} is performed in
the spirit of the standard multiscale method (interested readers
can consult, {\it
e.g.},~\cite{Bensoussan-Lions-Papanicolaou-1978,Majda-Kramer-1999}).
We consider the transport of a pollutant in a layer with
boundaries impermeable both for the fluid and for the pollutant.
In order to derive~\eref{eq1-06}, we substitute filtration
velocity $\vec{v}$ from~\eref{eq1-02} and pollutant flux $\vec{j}$
from~\eref{eq1-04} into conservation law~\eref{eq1-05}, and write
down
\begin{equation}
(\Psi_zC)_x-(\Psi_xC)_z-D(C_{xx}+C_{zz})=0
\label{app-01}
\end{equation}
\noindent
(the subscripts $x$ and $z$ indicate respective derivatives). The
absence of the fluxes of the pollutant and the fluid trough the
boundary, {\it \i.e.}, $j^z(z\!=\!0)=j^z(z\!=\!h)=0$ and
$v^z(z\!=\!0)=v^z(z\!=\!h)=0$ (the superscripts $x$ and $z$
indicate the respective components of vectors), results in
boundary conditions
\begin{equation}
z=0,\,h:\quad C_z=0\,.
\label{app-02}
\end{equation}

We assume $D\sim h^{-1}$, and use $h$ as a small parameter of
expansion;
 $\partial_z\propto h^{-1}\!$,
 $D=h^{-1}D_{-1}$, $f(z)=h^{-1}f_{-1}(z)$,
 $C(x,z)=C_0(x,y)+h\,C_1(x,y)+h^2C_2(x,y)+...$\,.
Then~\eref{app-01} reads
\begin{equation}
\begin{array}{l}
h^{-1}\partial_x[f_{-1,z}\theta_x(C_0+h\,C_1+...)]
 -h^{-1}\partial_z[f_{-1}\theta_{xx}(C_0+h\,C_1+...)]\\[8pt]
-h^{-1}D_{-1}[C_{0,xx}\!+C_{0,zz}\!+h(C_{1,xx}\!+C_{1,zz})
 +h^2(C_{2,xx}\!+C_{2,zz})\!+...]=0.
\end{array}
\label{app-03}
\end{equation}

From~\eref{app-03} \underline{in the order $h^{-3}$}:
\[
-C_{0,zz}=0\,.
\]
\noindent
Due to boundary conditions~\eref{app-02}, $C_{0,z}=0$, {\it i.e.},
in the leading order the concentration is uniform along $z$,
\[
C_0=\eta_0(x)\,.
\]

From~\eref{app-03} \underline{in the order $h^{-2}$}:
\[
\partial_x[f_{-1,z}\theta_x C_0]
 -\partial_z[f_{-1}\theta_{xx}C_0]
 -h\,D_{-1}C_{1,zz}=0\,,
\]
\noindent
{\it i.e.},
\[
C_{1,zz}=(h\,D_{-1})^{-1}f_{-1,z}(z)\,\theta_x(x)\,\eta_{0,x}(x)\,.
\]
\noindent
The last equation yields
\[
C_1=g_1(z)\eta_1(x)+A(x)z+B(x)\,,
\]
\noindent
where
\[
g_{1,zz}=h^{-1}f_{-1,z}=3\sqrt{35}\,(h-2z)/h^3,\quad
\eta_1(x)=D_{-1}^{-1}\,\theta_x(x)\,\eta_{0,x}(x)\,.
\]
\noindent
Formally, $g_1=3\sqrt{35}(z^2/2h^2-z^3/3h^3+az+b)$. Boundary
conditions~\eref{app-02} result in $A(x)=a=0$; $B(x)$ makes a
uniform along $z$ contribution to $C$, like $\eta_0(x)$, and,
therefore, can be treated as a part of $\eta_0$. Hence, we may
claim $\int_0^1C_1\rmd z=0$, {\it i.e.}, $B(x)=b=0$, and obtain
\[
C_1=g_1(z)\,\eta_1(x)
=3\sqrt{35}\left(\frac{z^2}{2h^2}-\frac{z^3}{3h^3}\right)\eta_1(x)\,.
\]

Let us now find the gross pollutant flux $h\,J$ through a vertical
cross-section of the layer. The integral of $j^x$
[equation~\eref{eq1-04}] over $z$ is
\begin{eqnarray}
 h\,J&\equiv&\int\limits_0^hj^x\rmd z=\int\limits_0^h(v^xC-DC_x)\rmd z
  =\int\limits_0^h(\Psi_zC-DC_x)\rmd z\nonumber\\
&=&\int\limits_0^h(f_z\theta_xC-DC_x)\rmd z
  =\int\limits_0^h(f_z\theta_x\eta_0+h\,f_zg_1\theta_x\eta_1-D\eta_{0,x}+O(h^3))\rmd z\nonumber\\
 &\approx&h\,\theta_x\eta_1\int\limits_0^hf_zg_1\,\rmd z-h\,D\eta_{0,x}
 =-h\left(\frac{21}{2h^2D}(\theta_x)^2\eta_{0,x}+D\eta_{0,x}\right).\nonumber
\end{eqnarray}
\noindent
As the pollutant is not accumulated anywhere, flux $J$ should be
constant along the layer. Hence, we find equation~\eref{eq1-06}
providing the relation between the concentration field and the
flux.


\section*{References}
\begin{thebibliography}{20}

\bibitem{Anderson-1958}
Anderson P W 1958
% {\em Absence of Diffusion in Certain Random Lattices,}
 \PR {\bf 109} 1492--505

\bibitem{Rossum-Nieuwenhuizen-1999}
van Rossum M C W and Nieuwenhuizen Th M 1999
% {\em Multiple scattering of classical waves: microscopy, mesoscopy, and diffusion,}
 \RMP {\bf 71} 313--71

\bibitem{Maynard-2001}
Maynard J D 2001
% {\em Colloquium: Acoustical analogs of condensed-matter
% problems,}
 \RMP {\bf 73} 401--17

\bibitem{Froehlich-Spencer-1984}
Fr\"ohlich J and Spencer T 1984
% {\em A rigorous approach to Anderson localization,}
{\it Phys.\ Rep.} {\bf 103} 9--25

\bibitem{Lifshitz-Gredeskul-Pastur-1988}
Lifshitz I M, Gredeskul S A and Pastur  L A 1988
 {\it Introduction to the Theory of Disordered Systems}
 (New York: Wiley).

\bibitem{Gredeskul-Kivshar-1992}
 Gredeskul S A and Kivshar Yu S 1992
% {\em Propagation and scattering of nonlinear waves in disordered systems,}
{\it Phys.\ Rep.} {\bf 216} 1--61

\bibitem{Pikovsky-Shepelyansky-2008}
 Pikovsky A S and Shepelyansky D L 2008
% {\em Destruction of Anderson Localization by a Weak
% Nonlinearity,}
 \PRL {\bf 100} 094101

\bibitem{Goldobin-Shklyaeva-PRE-2008}
 Goldobin D S and Shklyaeva E V 2008
 Localization and advectional spreading of convective flows under parametric disorder
 \PR E {\it submitted}
 [preview: arXiv:0804.3741]

\bibitem{Hammele-Schuler-Zimmermann-2006}
Hammele M, Schuler S and Zimmermann W 2006
% {\em Effects of parametric disorder on a stationary bifurcation,}
{\it Physica} D {\bf 218} 139--57

\bibitem{Knobloch-1990}
 Knobloch E 1990
% {\em Pattern selection in long-wavelength convection,}
 {\it Physica} D {\bf 41} 450--79

\bibitem{Shtilman-Sivashinsky-1991}
 Shtilman L and Sivashinsky G 1991
% {\em Hexagonal structure of large-scale Marangoni convection,}
 {\it Physica} D {\bf 52} 477--88

\bibitem{Aristov-Frick-1989}
 Aristov S N and Frik P G 1989
% {\em Large-scale turbulence in Rayleigh-B\'ernard convection,}
 {\it Fluid Dynamics} {\bf 24}(5) 690--5

\bibitem{Schoepf-Zimmermann-1989-1993}
 Sch\"opf W and Zimmermann W 1989
 {\it Europhys.\ Lett.} {\bf 8} 41--6

 Sch\"opf W and Zimmermann W 1993
 \PR E {\bf 47} 1739--64

\bibitem{Goldobin-Shklyaeva-BR-2008}
 Goldobin D S and Shklyaeva E V 2008
 Large-scale thermal convection in a horizontal porous layer
 \PR E {\it submitted}
 [preview: arXiv:0804.2825]

\bibitem{Michelson-1986}
 Michelson D 1986
 %{\em Steady Solutions of the Kuramoto-Sivashinsky Equation,}
 {\it Physica} D {\bf 19} 89--111

\bibitem{Goldobin-Lyubimov-2007}
 Goldobin D S and Lyubimov D V 2007
 %{\em Steady Solutions of the Kuramoto-Sivashinsky Equation,}
 {\it JETP} {\bf 104} 830--6

\bibitem{Frisch-1995}
 Frisch U 1995
 {\it Turbulence: The Legacy of A.\ N.\ Kolmogorov}
 (Cambridge: Cambridge University Press) p~226

\bibitem{Majda-Kramer-1999}
 Majda A J and Kramer P R 1999
%  Simplified models for turbulent diffusion:
%  Theory, numerical modelling, and physical phenomena
 {\it Phys.\ Rep.} {\bf 314} 238--574

\bibitem{Saffman-1960-Mazzino-Vergassola-1997}
 Saffman P G 1960
%  On the effect of the molecular diffusivity in turbulent diffusion
 {\it J. Fluid. Mech.} {\bf 8} 273--83

%\bibitem{Mazzino-Vergassola-1997}
 Mazzino A and Vergassola M 1997
%  Interference between turbulent and molecular diffusion
 {\it Europhys.\ Lett.} {\bf 37} 535--40

\bibitem{Bensoussan-Lions-Papanicolaou-1978}
 Bensoussan A, Lions J L and Papanicolaou G 1978
 {\it Asymptotic Analysis for Periodic Structures}
 (Amsterdam: North-Holland) %637 p

\end{thebibliography}
\end{document}